\documentclass[useAMS,usenatbib]{mn2e}
\usepackage{mn2e-breakabs}
\usepackage{rotating}       
\usepackage{graphicx}
\usepackage{amsmath} 
\usepackage{amssymb} 
\usepackage{natbib}
\usepackage{lscape}
\usepackage{color}

\def\kms{\mbox{km~s$^{-1}$}}

\def\kpch{\mbox{$h^{-1}$\,kpc}}

\def\LCDM{\mbox{$\Lambda$CDM}}

\def\mpch{\mbox{$h^{-1}$\,Mpc}}
\def\Mpch{\mbox{$h^{-1}$\,Mpc}}
\def\Mvir{\mbox{$M_{\rm vir}$}}
\def\M200{\mbox{$M_{\rm 200}$}}

\def\Msunh{\mbox{$h^{-1}M_\odot$}}

\def\R200{\mbox{$R_{\rm 200}$}}

\def\Vmax{\mbox{$V_{\rm max}$}}
\def\Vpeak{\mbox{$V_{\rm peak}$}}
\def\V200{\mbox{$V_{\rm 200}$}}

\title[HAM: conditions for convergence]{Halo Abundance Matching: accuracy and conditions for numerical convergence}

\author[Klypin et al.]{Anatoly\,Klypin$^{1}$\thanks{E-mail: aklypin@nmsu.edu},
  Francisco\,Prada$^{2,3,4}$, Gustavo\,Yepes$^{5}$,\newauthor
   Steffen\,He\ss$^{6}$, Stefan\,Gottl\"ober$^{6}$ \\
$^1$, Astronomy Department, New Mexico State University,
MSC 4500, P.O.Box 30001, Las Cruces, NM, 880003-8001, USA \\
$^2$ Instituto de F´ısica Te´orica, (UAM/CSIC), Universidad
Aut´onoma de Madrid, Cantoblanco, E-28049 Madrid, Spain \\
$^3$ Campus of International Excellence UAM+CSIC, Cantoblanco,
E-28049 Madrid, Spain\\
$^4$ Instituto de Astrof´ısica de Andaluc´ıa (CSIC), Glorieta de la
Astronom´ıa, E-18080 Granada, Spain\\
$^5$ Departamento de Fisica Teorica, Modulo C-15, Facultad de Ciencias, Universidad
Autonoma de Madrid, 28049 Cantoblanco, Madrid, Spain \\
$^6$ Leibniz-Institut f¨ur Astrophysik (AIP), An der Sternwarte 16,
D-14482 Potsdam, Germany
}

\begin{document}
\maketitle
\begin{abstract}
  Accurate predictions of the abundance and clustering of dark matter
  haloes play a key role in testing the standard cosmological model.
  Here, we investigate the accuracy of one of the leading methods of
  connecting the simulated dark matter haloes with observed galaxies
  -- the Halo Abundance Matching (HAM) technique. We show how to
  choose the optimal values of the mass and force resolution in
  large-volume $N$-body simulations so that they provide accurate
  estimates for correlation functions and circular velocities for
  haloes and their subhaloes -- crucial ingredients of the HAM
  method. At the 10\% accuracy, results converge for $\sim$50
  particles for haloes and $\sim$150 particles for progenitors
  of subhaloes. In order to achieve this level of accuracy a number of
  conditions should be satisfied. The force resolution for the
  smallest resolved (sub)haloes should be in the range $(0.1-0.3)r_s$,
  where $r_s$ is the scale radius of (sub)haloes. The number of
  particles for progenitors of subhaloes should be $\sim 150$. We
  also demonstrate that the two-body scattering plays a minor role for the
  accuracy of $N$-body simulations thanks to the relatively  small number of
  crossing-times of dark matter in haloes, and the limited force
  resolution of cosmological simulations.
\end{abstract}  


\section{Introduction}
\label{sec:intro}

The standard \LCDM~ cosmological model phases significant challenges
when it comes to testing its predictions for the distribution and
properties of galaxies. The model is able to make detailed predictions
on the distribution of dark matter, but connecting the luminous
galaxies with their dark matter haloes is a difficult task.  There are
different possibilities to make this galaxy-halo connection.  Halo
Abundance Matching (HAM) is a simple and yet realistic way to bridge
the gap between dark matter haloes and galaxies
\citep{KravtsovHOD04,Tasitsiomi2004,Vale04,conroy06,Guo10,
  Trujillo-Gomez,Reddick2012,Kravtsov2013}.  HAM resolves the issue of
connecting observed galaxies to simulated haloes and subhaloes by
setting a correspondence between the stellar and halo masses: more
luminous galaxies are assigned to more massive haloes.  By
construction, the method reproduces the observed stellar mass and
luminosity functions. It also reproduces the galaxy clustering over a
large range of scales and redshifts
\citep{conroy06,Guo10,Trujillo-Gomez,Reddick2012,Nuza2013}. When HAM
is applied using, for example, the observed SDSS stellar mass function
\citep{li09}, it gives also a reasonable fit to lensing results
\citep{Mandelbaum06}, to the galaxy clustering and the relation
between stellar and halo virial masses \citep{Guo10}.
\citet{Trujillo-Gomez} show that accounting for baryons drastically
improves the match of predicted and observed Tully-Fisher and Baryonic
Tully-Fisher relations. HAM modeling was also successful in
reproducing other properties of galaxies
\citep{Behroozi10,Leauthaud11,Reddick2012,Hearin2012,Kravtsov2013}.

Yet, using the HAM technique to connect galaxies to (sub)haloes
requires substantially more accurate $N$-body simulations as compared
with those used for the Halo Occupation Distribution (HOD) model
\citep{KravtsovHOD04,Tasitsiomi2004,Vale04}. Because of the demanding
numerical requirements, there are relatively few studies of abundance
and clustering of haloes based on circular velocities, which are often
used in conjunction with HAM models
\citep{Bolshoi,Trujillo-Gomez,Reddick2012,GuoWhite2013}.  Accuracy of
these statistics was challenged by \citet{GuoWhite2013}, who found a
very poor numerical convergence of results for subhalo population in
the Millennium I \citep{MSI} and Millennium II \citep{MSII}
simulations. Though we find much better convergence in our Bolshoi and
suite of MultiDark simulations, we agree with \citet{GuoWhite2013}
that it is important to investigate the accuracy and limits of HAM
technique to connect galaxies with (sub)haloes using only $N$-body
simulations.  In particular, physically robust results demand that
statistics such as the abundance and clustering of subhaloes to be
unaffected by numerical resolution.

Many large volume $N$-body cosmological simulations are not suited and
should not be used for HAM models. The numerical and physical
processes which affect the accuracy of results based on cosmological
simulations were discussed in many publications
\citep[e.g.,][]{Knebe2000,Klypin2001,Power2003,Hayashi2003}. It is one
of the goals of this paper to find conditions, that a simulation
should pass in order to be used for HAM models.

The key ingredient of HAM models are subhaloes: satellites (subhaloes)
of distinct haloes must be part of the abundance matching prescription
because each lump of dark matter with enough mass and concentration
should host a galaxy regardless whether that is the central object or
a satellite.  We call a halo distinct if its center is not inside the
virial sphere of even a larger halo. By definition, a subhalo is
always within the virial radius of another halo, which in this case is
called parent halo. In the sense of dynamical evolution subhaloes are
different objects because their physical properties can be
significantly affected (mostly through tidal stripping) by their
parent haloes \citep{Tasitsiomi2004,TumultuousLives}. In reality, the
boundary separating distinct haloes and subhaloes is blurry.  First,
there are different definitions of virial radius. As a result, the
same object can be called distinct or subhalo depending on the virial
ghalo definition. Second, distinct haloes may experience strong
interaction with their environment long before they formally cross the
virial radius of their parent \citep{TumultuousLives,Behroozi2013}. In
that respect some distinct haloes evolve as subhaloes.

There are different flavors of HAMs. Generally, one does not expect a
pure monotonic relation between stellar and halo masses. There
should be some degree of stochasticity in this relation due to
deviations in the halo merger history, angular momentum, and 
concentration. Even for haloes (or subhaloes) with the same mass,
these properties should be different for different systems, which
would lead to deviations in stellar mass. Observational errors are
also responsible in part for the non-monotonic relation between halo
and stellar masses.  Most of modern HAM models already implement
prescriptions to account for the stochasticity
\citep[e.g.,][]{Behroozi10,Trujillo-Gomez,Leauthaud11}.  The
difference between monotonic and stochastic models depends on the
magnitude of the scatter and on the stellar mass. The typical value of
the scatter in the $r$-band is expected to be $\Delta M_r =0.3$--$0.5$
mag \citep[e.g.,][]{Trujillo-Gomez}. For the Milky Way-size galaxies
the differences are practically negligible \citep{Behroozi10}.

Because haloes may experience tidal stripping, their dark matter mass
may be significantly reduced
\citep[e.g.][]{Klypin1999,Hayashi2003,TumultuousLives,Arraki}. Galaxies
hosted by these stripped haloes are not expected to lose their stellar
mass to the same degree as the dark matter mass because the stellar
component is much more concentrated and thus less susceptible to tidal
forces. This is the reason why the halo dark matter mass is expected
to be a poor indicator of the stellar mass of galaxy hosted by the
halo. There are quantities that should work better for HAM and are
often used as proxies for stellar mass: dark matter mass before the
stripping started, maximum circular velocity of the dark matter or the
maximum circular velocity of the dark matter before the stripping
\citep[e.g.,][]{conroy06,Trujillo-Gomez,Reddick2012}.
\begin{table*}
\centering
\begin{minipage}{175mm}
  \caption{Basic parameters of the cosmological
    simulations. $L_{box}$ is the side length of the simulation box,
    $N_p$ is the number of simulation particles, $\epsilon$ is the
    force resolution in comoving coordinates, $M_p$ refers to the
    mass of each simulation particle, and the parameters $\Omega_m$,
    $\Omega_{\Lambda}$ , $\Omega_b$, $n_s$ (the spectral index of the
    primordial power spectrum), $h$ (the Hubble constant at present
    in units of 100 $km/s Mpc^{-1}$) and $\sigma_8$ (the rms
    amplitude of linear mass fluctuations in spheres of 8 $h^{-1}Mpc$
    comoving radius at redshift $z=0$) are the $\Lambda$CDM
    cosmological parameters assumed in each simulation.}
 \begin{tabular}{@{}lccccclclccll@{}}
 \hline
  Name          &$L_{box}$& $N_p$& $\epsilon$& $M_p$ &$\Omega_m$
&$\Omega_{\Lambda}$ & $n_s$ & $h$ & $\sigma_8$ & code
&reference\\
                & ($\Mpch$) &               &  ($\kpch$) & 
($h^{-1}M_{\odot}$) &  &  &  &  &  & & & \\
\hline
Bolshoi         & 250  & $2048^3$ &  1 & $1.35\, 10^8$ & 0.27 & 0.73 &  0.95 & 0.70  & 0.82 & ART& Klypin+10 \\ 
MultiDark       & 1000 & $2048^3$  &   7  & $8.67\, 10^9$  &  0.27 & 0.73 & 0.95 & 0.70  & 0.82  &  ART  &Prada+12 \\
BigMultiDark    & 2500 & $3840^3$  &   10  & $2.06\, 10^{10}$  &  0.27 & 0.73  & 0.95 & 0.70  & 0.82  & LGadget-2 &Hess+13 \\

 \hline
Millennium      & 500  & $2160^3$ & 5  & $8.61\, 10^8$ & 0.25 & 0.75 &  1.00 & 0.73  & 0.90 & Gadget-2 &Springel+05   \\
Millennium-II   & 100  & $2160^3$ &  1 & $6.89\, 10^6$ & 0.25 & 0.75 &  1.00 & 0.73  & 0.90 & Gadget-3 &Boylan- \\
                &      &          &    &               &      &      &       &       &      &          &Kolchin+09 \\

\hline
\end{tabular}
\end{minipage}
\label{tab:Table1}
\end{table*}

Here we use the maximum circular velocity, not mass.  The maximum is
reached in the central halo region, which is expected to correlate
better with the stellar or luminous component and it is less sensitive
to tidal stripping. This is the main reason why in this paper we focus
our numerical convergence study on the abundance and clustering
statistics related with the (sub)halo circular velocities.

While the maximum circular velocity is a better quantity to characterize (sub)haloes
\citep{conroy06,Bolshoi,Trujillo-Gomez}, it is more difficult to accurately measure it in
numerical simulations. For example,  for galaxy-size haloes the maximum of the
circular velocity happens at $\sim 1/5$ of the virial radius. At this
radius, the resolution should be sufficient to estimate the maximum circular
velocity with better than, say, an accuracy better than 10 percent. In addition, subhaloes
also should be resolved and their circular velocities are
estimated. Thus, the resolution of simulations intended for HAM models
should be at least an order of magnitude better than those intended, for
example, for HOD models.

Our paper is organized as follows. In Section~\ref{sec:sims} we
present simulations used for our analysis and discuss halo
identification. Section~\ref{sec:Abundance} gives results on the
abundance of haloes and subhaloes. The numerical convergence study of
the correlation function for (sub)haloes selected by the maximum
circular velocity is discussed in Section~\ref{sec:Corr}. In
Section~\ref{sec:Estimate} we investigate numerical and physical
processes, which affect accuracy of cosmological $N$-body
simultions. Here we also provide the conditions for the numerical
convergence of HAM results.  Our results are summarized in
Section~\ref{sec:Conclusions}

\section{Simulations and dark matter Haloes}
\label{sec:sims}

In order to study the numerical convergence of cosmological $N$-body
simulations we use three large simulations, which cover a mass range
of five orders of magnitude. The same cosmological parameters were
used for the simulations. The simulations were performed using two
codes: the Adaptive Refinement Tree \citep[ART][]{ART1997,ART2008} and
Gadget \citep{2005MNRAS.364.1105S}. Details of the ART simulations
Bolshoi and MultiDark were given in \citet{Bolshoi} and
\citet{Prada2012}.  The simulation done with the Gadget  is from a
large suit of BigMultiDark simulations\citep{Hess2013}.
Main parameters of our simulations are presented in Table~\ref{tab:Table1},
where we also present parameters of the two Millennium
simulations \citep{MSI,MSII}.  The table highlights one substantial
difference between our simulations and  Millennium: our
force resolution is substantially better for the same mass
resolution. Another difference, which is not shown in the table is that
we systematically use significantly smaller time-steps.

The choice of numerical parameters (combination of mass, force, and
time resolutions) in our simulations is not accidental. We made many
tests (some of them discussed below) studying the convergence of results.

\begin{figure}
      \includegraphics[width=84mm]{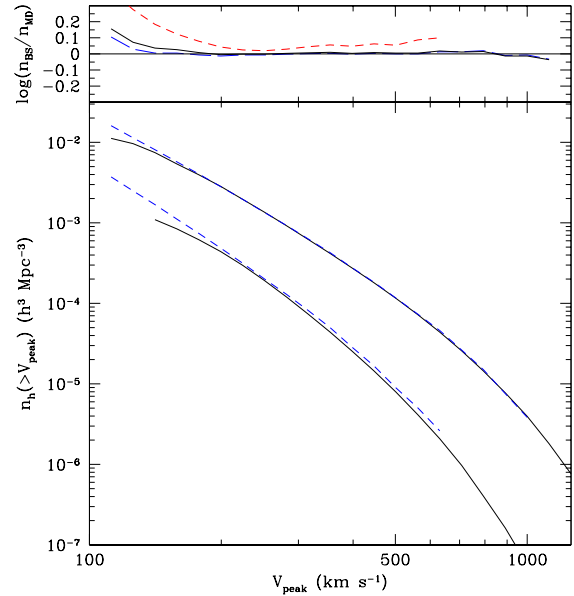}
      \caption{Cumulative velocity function of dark matter haloes at
        $z=0$ for Bolshoi and Multidark simulations. Here we use the
        maximum circular velocity $V_{\rm peak}$ over the merging
        history of each halo. {\it Bottom panel:} Top curves show the
        total number of haloes (distinct + subhaloes). Bottom curves
        show subhaloes only. Full (dashed) curves are for the Multidark
        (Bolshoi) simulation. {\it Top panel:} the ratio of the number
        of haloes in Bolshoi and Multidark simulations.  The full
        curves show all haloes. The long and short dash curves show
        distinct and subhaloes respectively.  At the 10\% level the total number of
        haloes converge for $V_{\rm peak}>140\,\kms$.  Convergence for
        subhaloes is reached for larger velocities $V_{\rm
          peak}=200\,\kms$, corresponding to 150 particles in the
        MultiDark simulation.}
\label{fig:Vacc}
\end{figure}

We use the Bound Density Maxima (BDM) spherical overdensity code
\citep{1997astro.ph.12217K,Riebe11} to identify and find the properties
of haloes and subhaloes in all our simulations. The code was extensively
tested and compared with other halofinders
\citep{Knebe11,RockStar,Knebe2013}. Halo tracking was made for two
simulations (Bolshoi and MultiDark) using the ROCKSTAR code
\citep{RockStar}. No halo tracking so far is available for the bigMultiDark
simulation. 

\begin{figure*}
      \includegraphics[width=0.9\textwidth]{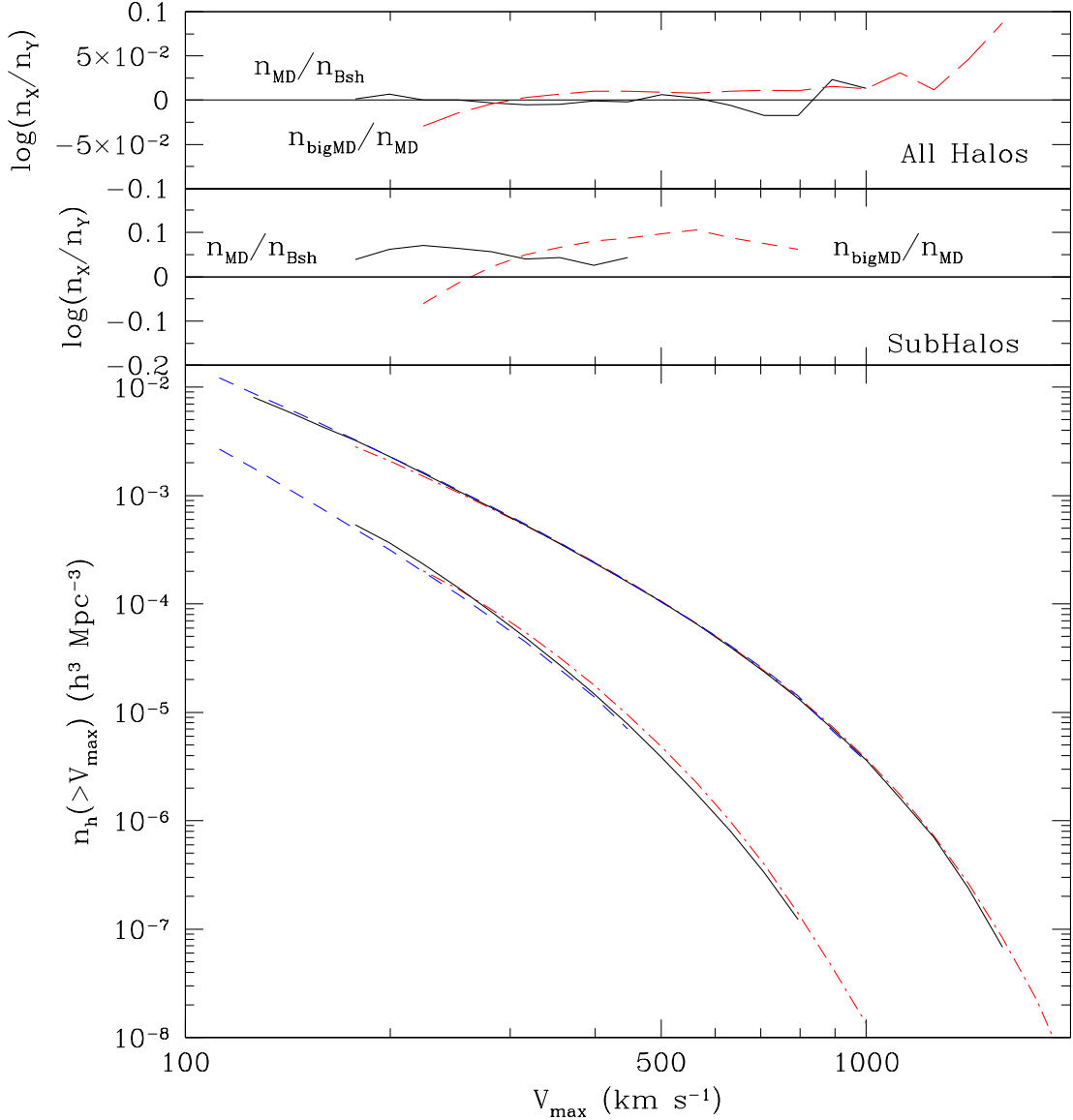}
      \caption{Comparison of velocity functions in simulations with
        vastly different resolutions. {\it Bottom panel:} Top curves
        are total number of haloes. Lower curves show only
        subhaloes. Dashed, full, and dot-dashed curves are for
        Bolshoi, MultiDark and BigMultiDark respectively. Two top
        panels show the ratios of all haloes and subhaloes for
        different pairs of simulations as labeled in the plot.  }
\label{fig:Vmax}
\end{figure*}

The BDM halo finder provides many parameters for each (sub)halo including 
the  virial mass, concentration, and spin parameter. It also
includes the maximum circular velocity $\Vmax = \sqrt{GM(r)/r}\vert_{\rm
  max}$. For Bolshoi and MultiDark  we also find the peak
of $\Vmax$ over the history of evolution of each halo and subhalo. This is done using
the information over the history of the major progenitor of each
halo. The peak velocity $V_{\rm peak}$ serves as proxy for
the rotation velocity of the galaxy hosted by its halo. This is why $V_{\rm peak}$ plays a
key role for HAM models. However, from the point of view of the
accuracy of simulations, $\Vmax$ is a more important quantity with
$V_{\rm peak}$ simply considered to be  mapping $\Vmax$.

The virial radius for distinct haloes is defined as radius within which the
mean matter density is equal to the virial overdensity found from
the solution of the top-hat model for cosmological model with a
cosmological constant. For our set of cosmological parameters this
corresponds to a matter overdensity of 360 times the average matter
density in the universe.

When studying convergence of numerical results, we quote the number of
particles within the virial radius. Subhalos normally do not extend to
virial radius because their radius is truncated by tidal stripping.
In order to make the number of particles in haloes and subhalos
compatible, for subhalos we quote the number of particles inside the
virial radius of a distinct halo with the same maximum circular
velocity $\Vmax$ as that of the subhalo. The extrapolation to the
virial radius depends on halo concentration. We use the average virial
mass - $\Vmax$ relation for distinct halos to make the extrapolation
to the virial radius. The actual number of particles in a subhalo can
only be smaller than the extrapolated value. We find that the
extrapolated virial masses of subhalos are within 10--20\% of virial
masses of 90\% of progenitors of the subhalos defined at the moment of
the peak of $\Vmax$. This is why we call these estimates of the
subhalo masses and number of particles the masses and particles
numbers of subhalo progenitors.

\section{Halo velocity function}
\label{sec:Abundance}

We start with the analysis of the velocity functions using the peak
values of $\Vmax$ in Bolshoi and MultiDark. Figure~\ref{fig:Vacc}
shows the velocity functions separately for subhaloes and the total
populations of haloes.  Overall there is an excellent agreement
between both simulations for a wide range of circular velocities
$V_{\rm peak}= 200-1000\,\kms$, where (sub)haloes are resolved with
enough particles and the cosmic variance does not effect the results.
At large velocities ($V_{\rm peak}> 1000\,\kms$ there are clearly some
effects due to the lack of long waves and cosmic variance in the
Bolshoi simulation due to its smaller box size.  Deviations of the
number of haloes at small velocities $V_{\rm peak}<200\,\kms$ are
indication of numerical resolution effects in the MultiDark
simulation. The smaller number of haloes in Multidark at smaller
velocities are due to the lack of force and mass resolution as
compared to Bolshoi (see Table~1).

At the 10\% level of accuracy the number-density of haloes converge for
peak velocities $V_{\rm peak}>140\,\kms$, which corresponds to the
average number of 52 particles inside the virial radius in the
Multidark simulation.  As expected, convergence is worse for subhaloes.
However, it is still very good: at 10\% it is $V_{\rm
  peak}=200\,\kms$. For isolated haloes this corresponds to an average
of 150 particles. This is a remarkably small number of particles. For
comparison, convergence of the velocity function of subhaloes in
the Millennium simulations, at the same accuracy level, was achieved only
with $\sim 3000$ particles \citep[see][]{GuoWhite2013}

Accuracy of HAM results depends on a number of numerical effects. Each
of them should produce small enough errors so that the final
result can be trustful.  The effects include the tracking of subhaloes
as they fall into larger haloes and become subhaloes \citep{Behroozi10}. Other
effects include mass and force resolution, time-integration of
equations of motion, and two-body scattering. Accuracy of halo
tracking was discussed in \cite{Behroozi10}.  Here we focus on the other
effects. In order to study those, we investigate the statistics based on
the instantaneous maximum of the circular velocity $\Vmax$.

Figure~\ref{fig:Vmax} presents the results of three simulations: Bolshoi,
MultiDark, and BigMultiDark. To large degree the results are very
similar to those based on $V_{\rm peak}$. Again, the comparison between
 Bolshoi and MultiDark shows convergence at 10\% level for
distinct haloes with $\Vmax>140\, \kms$ and for subhaloes with $\Vmax>200\,
\kms$.  This implies that subhaloes need about 150 particles to have
this accuracy. The comparison between BigMultiDark and MultiDark
shows the same, i.e. in order to reach the 10\% accuracy, one needs $\sim
150$ particles for subhaloes.

\section{Convergence of the correlation function}
\label{sec:Corr}

\begin{figure}
  \includegraphics[width=84mm]{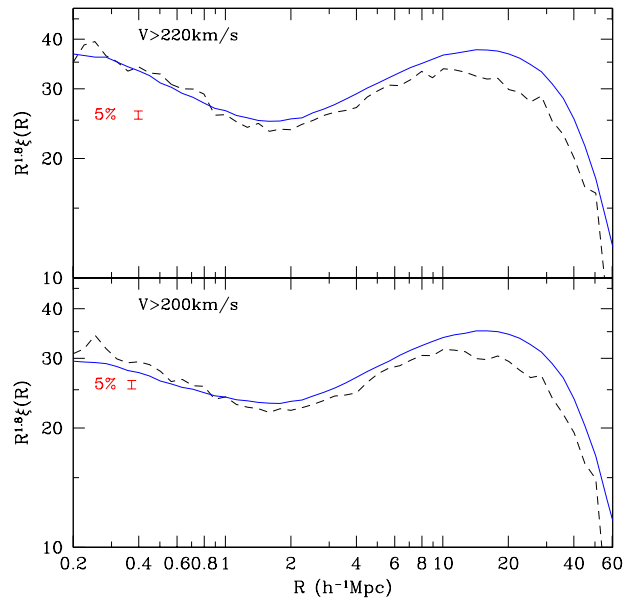}
  \caption{Comparison of the correlation functions for haloes in the
    MultiDark (full) and Bolshoi (dashed) simulations. Haloes were
    selected to have their peak circular velocities (maximum circular
    velocity over the accretion history of each halo) to be larger
    than $\Vpeak>200\,\kms$ for the bottom panel and
    $\Vpeak>220\,\kms$ for the top panel. Because of the lack of long
    waves in the Bolshoi, the correlation function in the MultiDark is
    larger for scales above $2\mpch$. Clustering at smaller scales is
    not affected by long waves and it is a test of numerical effects
    and convergence of simulations. At the 5\% level the correlation
    functions agree for both simulations. For the MultiDark
    simulation, velocity limit $\Vpeak>200\,\kms$ corresponds on
    average of 160 particles.}
\label{fig:corrVV}
\end{figure}

\begin{figure}
      \includegraphics[width=84mm]{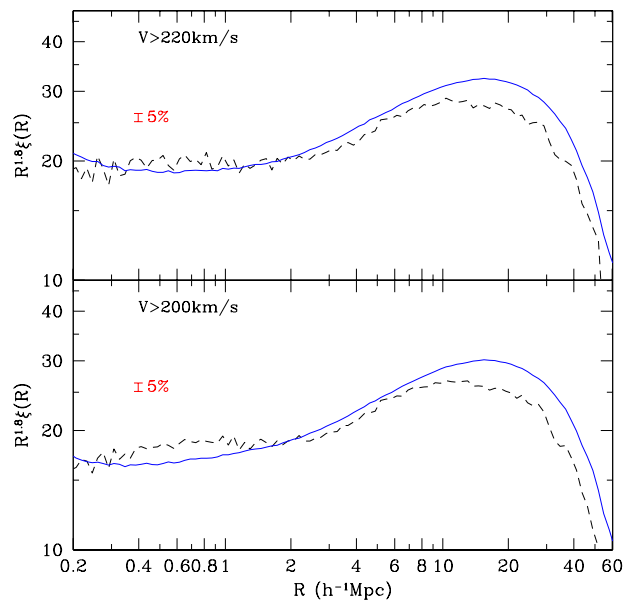}
      \caption{The same as in Figure~\ref{fig:corrVV}, but for haloes
        selected by the maximum circular velocity $\Vmax$. Clustering
        at scales below $2\mpch$ indicate convergence of results at
        $\sim 5\%$ level for haloes with more than $\sim 150$ particles.}
\label{fig:corrC}
\end{figure}

Convergence of the two-point correlation function for haloes and
subhaloes, selected by either $\Vpeak$ or $\Vmax$, is demonstrated in
Figures~\ref{fig:corrVV} and \ref{fig:corrC}, where we compare results
for Bolshoi and MultiDark. Just as for the circular velocity
functions, the agreement of the results on small scales ($\lesssim
2\Mpch$) indicates convergence of the results regaring the numerical
parameters such as mass and force resolution. The differences in
correlation functions on larger scales are due to finite box size and
cosmic variance.

On scales $0.2\Mpch<R<1\Mpch$ the agreement between Bolshoi and MultiDark
 is about 5\%, with  Bolshoi exhibiting slightly larger
clustering signal as expected due to it better force and mass
resolution (see Table~1). The situation is unclear on even smaller
scales mostly because of very small number of pairs within these
separations (and thus large noise) in Bolshoi. 

\begin{figure}
      \includegraphics[width=84mm]{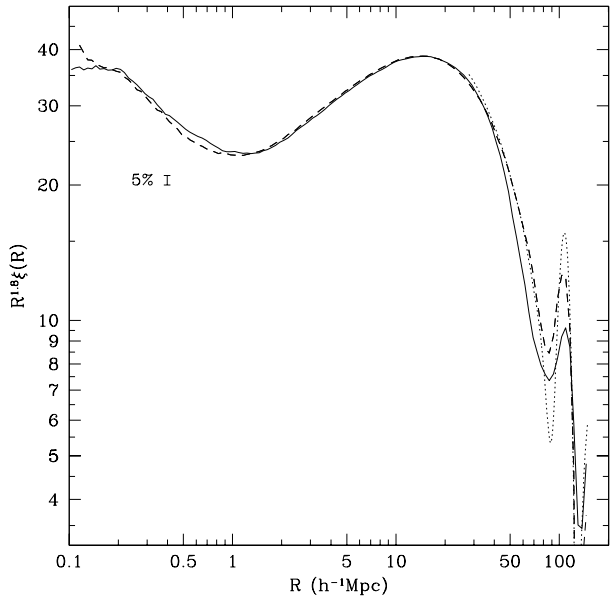}
      \caption{Correlation functions of haloes in the MultiDark (full)
        and BigMultiDark (dashed) simulations. Haloes were selected to
        have the maximum circular velocity $\Vmax>240\,\kms$, which on average
        corresponds to 120 particles. Both simulations have nearly the same (within few
        percent) correlation functions from 100\,\kpch~ to
        40\,\mpch. At larger distances the correlation function in the
        MultiDark is below that of BigMultiDark because of the cosmic
        variance. The dot-dashed curve represents the linear
        correlation function scaled up with bias factor 1.3. It shows
        that on scales larger than $\sim 30\mpch$~ clustering of
        haloes has a nearly scale-independent bias. The correlation
        function around the BAO peak is slightly damped and broadened
        by non-linear effects.  }
\label{fig:corrD}
\end{figure}

Because of much larger statistics of halos in MultiDark and
BigMultiDark simulations, the level of shot noise on $\sim 100\,\kpch$
scales is smaller, which allows us to test those simulations on
these scales.  Indeed, in this case the converge of results is
substantially better as illustrated by
Figure~\ref{fig:corrD}. It is also much better on large scales due to
sugnificantly larger box sizes.  Both simulations have nearly the same
(within few percent) correlation functions from $\sim 100\,\kpch$~ to
$40\,\mpch$.  One starts to run on substantially larger differences on
scales below $\sim 100\,\kpch$. This is due to a combination of force
resolution and real-space halofinder, which misses subhalos, when
their centers are too close to the center of parent halo.

\begin{figure}
      \includegraphics[width=84mm]{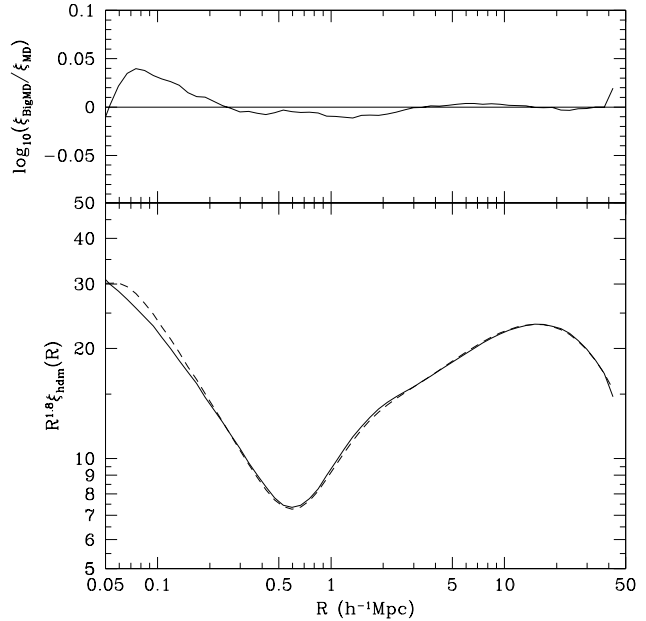}
      \caption{{\it Bottom panel:} Cross correlation function of halos
        and dark matter in MultiDark (full) and BigMultiDark (dash)
        simulations.  {\it Top panel:} Ratio of the cross
        correlation functions. Distinct halos with $\Vmax
        =(250-270)\kms$ were used. }
\label{fig:Cross}
\end{figure}

Another interesting statistics, which can be used to probe the
numerical convergence, is the halo-dark matter cross correlation
function $\xi_{\rm hdm}$. It is an important quantity on its own
because it is required for theoretical predictions of the weak galaxy
lensing signal \citep[e.g.,][]{Leauthaud2011}.  Figure~\ref{fig:Cross}
shows $\xi_{\rm hdm}$ for MultiDark and BigMultiDark with halos
selected at the low limit of resolution: $\Vmax =(250-270)\kms$. The
difference between the simulations is less than $\sim 2$\% for scales
larger than $\sim 150\,\kpch$. There is a 5--10\% error on scales
$\lesssim 100\,\kpch$ with the lower-resolution BigMultiDark
simulation predicting stronger clustering signal. By using simple
$N$-body toy models we find that the reason for this extra clustering
is due to a side-effect of force resolution. Because of lack of
resolution, the dark matter, which should have been at small
distances, ends up at larger distances, where it slightly increases
the local density and, thus, increases the cross correlation signal.

\begin{figure}
      \includegraphics[width=84mm]{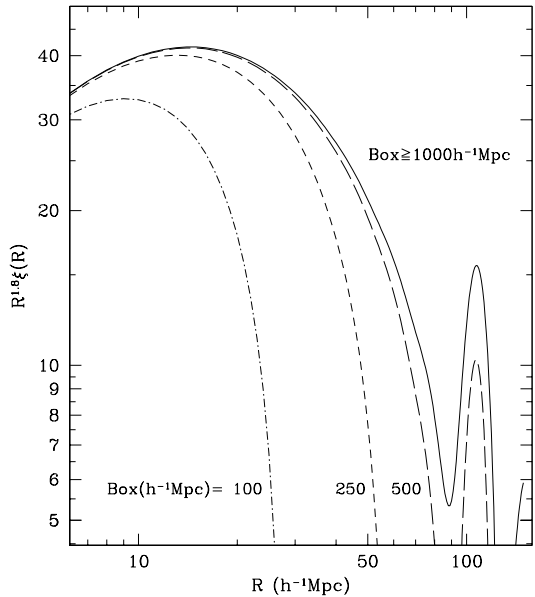}
      \caption{Effects of the finite simulation box size on the
        correlation function. We use the linear power spectrum scaled
        up by the bias factor $b=1.7$ to calculate the correlation
        functions. In order to mimic the lack of long waves in
        simulations of different box sizes, the integral over the
        power spectrum is truncated at $k_{\rm Box} =2\pi/L$, where
        $L$ is the size of the computational box. The lack of
        long waves in small boxes results  in significant
        suppression of clustering at scales even as small as 1/10 of
        the computational box.}
\label{fig:corrL}
\end{figure}

On scales above $2\Mpch$ the correlation function in Bolshoi is
systematically lower than in MultiDark.  The large-scale differences
between Bolshoi and Multidark simulations are likely due to both the
cosmic variance and the finite box size.  It is easy to test the
effect of box size.  For that, we estimate the correlation function
using linear theory and truncate the power spectrum at the lower
wavenumber limit given by the fundamental mode of the simulation box:
$k_{\rm Box}=2\pi/L$, where $L$ is the length of computational box.
Figure~\ref{fig:corrL} shows correlation functions of the dark matter,
which are obtained for different box sizes.

As expected, the finite size of the simulation box affects the large
scales in a profound way. For example, for the $L=100\,\Mpch$ box used
in the Millinium-II simulation \citep{MSII}, the correlation function
becomes {\it negative} for $R>25\Mpch$ while it should be
positive. Even smaller scales are affected: at $R=10\,\Mpch$ the
correlation function is 30\% below the true one (solid cirve in
Figure~\ref{fig:corrL}. Note that often used a rule of thumb that
scales smaller than 1/10 of the box length are not affected by the box size,
does not really hold. Indeed, the deviations of the linear correlation function at
$R=L/10$ strongly depend on the box length $L$. One may expect that nonlinear effects
exacerbate the differences.

Because of the box size effects, the comparison between simulations on
large scales should be done using a large box size. Indeed, the
converge of results is substantially better for large-box
simultions. Figure~\ref{fig:corrD} illustrates this point by comparing
MultiDark and BigMultiDark simulations with $L=1\,h^{-1}$Gpc and
$L=2.5\,h^{-1}$Gpc. Both simulations have nearly the same (within few
percent) correlation functions from $100\,\kpch$~ to $40\,\mpch$ for
haloes and progenitors of subhaloes with $\sim 150$ particles.

Convergence and accuracy of halo clustering and abundances in our set
of simulations are significantly better -- by a factor of 20 in mass
-- than for the Millennium simulations \citep{GuoWhite2013}. As a way
to compensate the lack of resolution in the Millennium simulations, a
trick is used: if a subhalo is lost, it is assigned to the most bound
dark matter particle of this subhalo, and called an ``orphan''. A
simplified dynamical friction estimate is then used to set a clock
when the subhalo is assumed to merge with its parent
\citep{Guo2011}. It is not clear to us what causes such a poor
performance of the Millennium simulations. This is not related with
the $N$-body code: we achieve satisfactory results aslo with the
Gadget code. There are some concerns regarding the Millennium
simulations, but none of them sepearately seems to be enough to
explain the poor convergence in the Millennium simulations. For
example, the number of time-steps is low. For the Millennium-I run the
parameter {\it
  ErrTolIntAccuracy}$=0.01$, which defines the number of steps, is twice larger
than what we use in our Gadget simulation bigMultidark.  The force
resolution is also low in the Millennium simulations. In addition, the
small size of the computational box of the Millennium-II run raises
doubts that it was large enough to provide the true solution when
compared with the Millennium-I. It is possible that a combination of
all these small defects resulted in much worse results for
subhaloes. We highlight that the better accuracy of our simulations
eliminates the need for ``orphans'' in our simulations.

\section{Conditions for convergence}
\label{sec:Estimate}
Usually, numerical convergence of cosmological simulations is
discussed in the context of simulations of individual dark matter
haloes with large number of particles.  For example, convergence of
dark matter density profiles is studied by running simulations with
increasing very large number of particles
\citep[e.g.][]{Klypin2001,Aquarius,Stadel2009}. Here, we are
interested in convergence on the opposite side of the spectrum: how
reliable are results with very small number of particles? The small
haloes are very important. After all, most of haloes in each
simulation are tiny, and numerous statistics (e.g.  the halo mass
functions) use those small halos. Results presented in
Section~\ref{sec:Abundance} indicate convergence of the velocity
function of distinct haloes with $\sim 50$ particles. How can we
understand this?

The accuracy and convergence of results used for halo abundance
matching depend on a number of physical and numerical processes. We
split those into two categories: (a) accuracy of the maximum circular
velocity $\Vmax$ for isolated haloes and (b) physics and numerics of
subhaloes.  We focus on some rather specific problems, which to large
degree define convergence of numerical results. The goal is to study
these processes separately by isolating each one and investigating it
by using a simple realistic model.  In reality, all processes are
interconnected and only a full-scale cosmological simulation have all
all of them included. However, understanding the conditions for
convergence of large cosmological simulations is difficult because it
is not clear what are the processes and how they potentially affect the
results. We get much better insights by studying separate processes.

\subsection{Accuracy of circular velocities of isolated haloes}
Here we consider three effects: (a) force resolution, (b) mass
resolution (number of particles) and (c) two-body scattering.  We
investigate a series of simulations of isolated NFW halos. In all
cases we start with a NFW halo, which is set in equilibrium.  The
equilibrium initial conditions are constructed assuming an ideal
phase-space distribution function for a NFW halo with isotropic
velocities and with no correction for force softening or
discreteness. Because of different numerical effects -- the force
softening, discreteness effects, and time-stepping -- the system of
$N$ particles does not stay in the equilibruim. It evolves. In most of
the cases, the system reaches another equilibrium after few dynamical
times. By comparing this new state with the initial one, we estimate
the magnitude of the numerical effect responsible for the evolution.

When dealing with the NFW halo, it is convenient to express all
physical quantities using the scale radius $r_s$ and the maximum
circular velocity $\Vmax$. Those quantities are related to the virial
mass $\Mvir$ and concentration $c$ through the following relations:
\begin{eqnarray}
   V^2_{\rm max} &=& \frac{G\Mvir}{R_{\rm vir}}\frac{f(x_{\rm max})}{f(c)}\frac{c}{x_{\rm max}},
    \label{eq:vmax} \\
   c &=& \frac{R_{\rm vir}}{r_s},\quad x\equiv \frac{r}{r_s}, \quad x_{\rm max}=2.163 \\
   M(x) &=& \Mvir\frac{f(x)}{f(c)}, \quad  f(x) = \ln(1+x) -\frac{x}{1+x},\label{eq:Mnfw}\\
   V^2(x) &=& V^2_{\rm max} \frac{x_{\rm max}}{x}\frac{f(x)}{f(x_{\rm max})} \label{eq:Vnfw}\\ 
   \rho(x) &=& \frac{V^2_{\rm max}}{4\pi Gr^2_s}\frac{x_{\rm max}}{f(x_{\rm max})}\frac{1}{x(1+x)^2}.
    \label{eq:DenNFW}
\end{eqnarray}
Written in this way, the halo structure
 does not depend on $c$ and
$\Mvir$. This allows us to simulate just one system and then to
rescale it to any particular values of $\Mvir$ and $c$. 

It is convenient to use some fiducial virial radius and virial
mass. For this, we use concentration $c=8$ and fiducial virial radius
$8\, r_s$ which are typical parameters for galaxy-size haloes with mass $M_{\rm
  vir}=10^{12}\Msunh$ \citep{Prada2012}. A useful scale for time is
the crossing-time at the radius $r_{\rm max}=2.163\, r_s$, at which the
circular velocity reaches the maximum $\Vmax$:
\begin{equation}
  t_{\rm cross} \equiv \frac{r_{\rm max}}{\sigma_v(r_{\rm max})}\approx x_{\rm max}\frac{r_s}{\Vmax},
\label{eq:tcross}
\end{equation}
where $\sigma_v(r_{\rm max})$ is the 3D {\it r.m.s.} velocity at $r_{\rm
  max}$. The crossing-time in physical units depends on halo
concentration. For virial mass defined at overdensity
$200\,\rho_{\rm cr}$, where $\rho_{\rm cr}$ being the critical density
of the Universe, the crossing-time is:
\begin{equation}
  t_{\rm cross} = 9.8\,h^{-1}\frac{f(c)}{c^{3/2}}{\rm Gyrs}.
\label{eq:cross}
\end{equation}
He have $t_{\rm cross}=0.3$~Gyr for a Milky Way-type galaxy with
$c=8$. The crossing-time depends only very weakly on halo mass: it is
twice smaller for haloes of dwarf galaxies with $M_{\rm
  vir}=10^{9}\Msunh$ and twice larger for clusters of galaxies with
$M_{\rm vir}=10^{15}\Msunh$.  Note that haloes do not live long:
during 10~Gyrs of evolution they have only $\sim 20-60$
crossing-times.

Initial density profiles extend to $\sim 50\, r_s$, which is
significantly larger than any realistic halo in cosmological
simulations.  All simulations use a simple direct summation code with
a leap-frog integration scheme. The code adopts the Plummer softening
(force resolution) $\epsilon$. All simulations were done with
time-steps small enough to have energy conservation $|\Delta E/E|<
10^{-6}$.

As a test, we run a simulation with large number of particles
$N=10,000$ and small force softening $\epsilon=0.005\, r_s$. With
the exception of very small non-equilibrium effects at the center, the
halo did not change for many dynamical time-scales.  However, when
simulations are run with either small $N$ or large $\epsilon$, the
haloes do get modified. The comparison of the evolved halo with its
initial equilibrium distribution gives as a way to measure effects of
mass and force resolution separately. In cosmological simulations
haloes grow in a complex fashion through merging and
accretion. However, at late stages of evolution the accretion slows
down and haloes spend most of their time in the regime of slowly
growing quasi equilibrium NFW distribution. Ability of code to
maintain this equilibrium distribution is an important indication of
its accuracy.

\subsubsection{Force resolution}
\label{Sec:Force}
We first study the effects of force resolution by 
studying  relatively large numerical experiments with 50,000 particles
within a radius of $10\, r_s$ and 100,000 particles in
total. Within the radius $x_{\rm max}=2.16\, r_s$ the halo has about 15,000
particles. This configuration was simulated with different force
resolutions ranging from $\epsilon =0.005\, r_s$ to $\epsilon
=2\, r_s$. Each simulation started with exactly the same initial
conditions of equilibrium NFW halo without any consideration for the
force resolution. For simulations with small $\epsilon$ there was very
little evolution in the distribution of particles.  However, in the cases
of large $\epsilon$ the force was too weak in the central region to
keep particles in equilibrium. As the result, each simulation with
large $\epsilon$ evolved starting with the center. All simulations
with $\epsilon <r_s$ settled into new equilibrium after few crossing-times.
  The simulation with $\epsilon=2\, r_s$ was changing even after 5
crossing-times. At this moment the simulation was stopped because its
very peripheral regions with $r>10\, r_s$ were affected.

\begin{figure}
      \includegraphics[width=84mm]{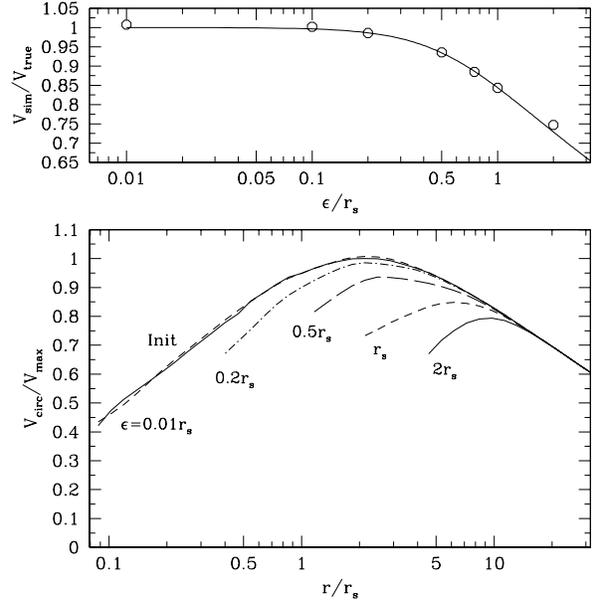}
      \caption{Effects of the force resolution on the structure of
        dark matter haloes.  {\it Bottom panel:} Circular velocity
        profiles of haloes with different force resolutions
        $\epsilon$. The top dashed curve shows the initial NFW
        profile. Each profile is displayed starting with radius
        $2\epsilon$. Profiles converge to the NFW profile as
        $\epsilon$ decreases. For each $\epsilon$ the profile is
        similar to NFW with larger effective scale radius and smaller
        maximum circular velocity. {\it Top panel:} dependence of the
        maximum circular velocity on the force resolution. The full
        curve shows approximation equation~(\ref{eq:Vcorrect}). }
\label{fig:ForceResolution}
\end{figure}

Figure~\ref{fig:ForceResolution} shows circular velocity profiles for
runs with different force resolutions $\epsilon$. Qualitatively
the results are similar to those found in convergence studies of
individual haloes in cosmological simulations: as the resolution
increases, halo circular velocity profiles converge.

We can use these results to estimate errors in $\Vmax$. The top panel
in Figure~\ref{fig:ForceResolution} shows the dependance of $\Vmax$ on
$\epsilon$.  At small $\epsilon$ the error in $\Vmax$ is relatively
small. For example, we have $\Vmax =0.98\,\Vmax_{,\rm true}$ for
$\epsilon= r_s/4$. However, the error of $\Vmax$ significantly
increases with increasing $\epsilon$: $\Delta\Vmax/\Vmax = 0.17$ for
$\epsilon =r_s$.  One must keep in mind that $\Vmax$ should be quite
accurate because the relevant statistics, which use $\Vmax$, are very
sensitive to it.  For example, the abundance of haloes and subhaloes
scales with the maximum circular velocity as $n(>\Vmax)\propto
\Vmax^{-3}$.

It is interesting to compare the convergence results in these
experiments with those in our cosmological simulations. Our results
indicated the convergence of the halo velocity function at the 5\%
accuracy for haloes with $\Vmax=240\,\kms$ in the BigMultiDark
simulation. This corresponds to haloes with 120 particles and typical
scale radius $r_s=30\,\kpch$. The force resolution is $\epsilon=7\,\kpch$ for the
BigMultiDark. This gives $\epsilon =0.23\,r_s$.
According to Figure~\ref{fig:ForceResolution} 
the error in $\Vmax$ is less than 2\%. In turn, the error in the velocity
function $n(>\Vmax)$ should be less than 5\%, which is close to what
we find in our cosmological simulations.

The following model can be used to make an approximation for the
errors in $\Vmax$. We note that profiles of $\Vmax(r;\epsilon)$ shown
in the bottom panel of Figure~\ref{fig:ForceResolution} look similar
to the NFW profile, which scale radius is larger and $\Vmax$ is
smaller than in the initial NFW halo.  At large radii all the velocity
curves converge to the same values  because the mass
distribution at large distances is not affected by the resolution. The
increase in the scale radius due to the force resolution can be approximated as
\begin{equation}
  r_{s, \epsilon} = \beta(\epsilon)r_s, \quad \beta(\epsilon) 
                  \approx \sqrt{1+(1.7\epsilon/r_s)^2}.
\label{eq:rs}
\end{equation}
Using this expression and normalizing the modified NFW profile so that
at large radii it has the same mass as the initial NFW,
one obtains an approximation for the decline of $\Vmax$ due to the
force resolution:

\begin{equation}
  \left(\frac{V_{\rm max, \epsilon}}{\Vmax}\right) \approx\beta(\epsilon)^{-1/4}
   =\left(1+(1.7\epsilon/r_s)^2\right)^{-1/4}.
\label{eq:Vcorrect}
\end{equation}
 This approximation is shown in the top
panel of Figure~\ref{fig:ForceResolution}.

The increase in the scale radius due to poor force resolution leads to
a decrease in halo concentration. Equation~(\ref{eq:rs}) implies that
concentration $c_\epsilon$ measured for a halo, which was simulated
with the force resolution $\epsilon$, declines as $ c_\epsilon
=c/\beta(\epsilon)$. For $\epsilon/r_s= 0.25$ this gives 8\% error in
concentration. If equation~(\ref{eq:vmax}) is used to find halo
concentration from measured $\Vmax$ and virial velocity $V^2_{\rm
  vir}=G\Mvir/R_{\rm vir}$ \citep[e.g.][]{Prada2012}, then reqirements
to the force resolution are even more stringent. Error analysis shows
that in order to have an error in concentration less than 5\%, the
force softening must be smaller than $\epsilon/r_s= 0.1$, which in
turn implies not more than 1\% error in $\Vmax$.

Equation~(\ref{eq:rs}) can be used to recover
the true concentration $c = R_{\rm vir}/r_s$. We assume that a
simulation, which is performed with the force resolution $\epsilon$,
provides the virial radius $R_{\rm vir}$ and the concentration $c_\epsilon$
for a  halo. The true concentration for the halo can be estimaed as:
\begin{equation}  
  c  = \frac{c_\epsilon}{\sqrt{1-(1.7\epsilon c_\epsilon/R_{\rm vir})^2}}. 
\label{eq:Conc}
\end{equation}

These approximations can be used to set the upper limit on the force
resolution needed to achieve a given accuracy of the maximum of the
circular velocity $\Vmax$. For example, in order to have no more than
2(5)\% error in $\Vmax$, the force resolution should be smaller than
$\epsilon/r_s< 0.25 (0.40)$. 

\subsubsection{Mass resolution}
Here we address two issues: (a) are there any systematic effects that
force the value of $\Vmax$, measured in simulations with small number
of particles, to deviate from its true value and (b) what is the level
of statistical fluctuations of $\Vmax$.

In order to address these issues, we made three simulations of NFW
halos, which were initially in equilibrium, each with 300 particles
inside radius of $52\, r_s$. These simulations initially had only
150 particles within the fiducial virial radius and 25 particles
inside $r_s$. This is the smallest number of particles, for which our
cosmological simulations show convergence.  Results presented below
are averages of the three realizations. The force resolution is
$\epsilon=0.2\, r_s$, which is typical for our cosmological
simulations for the smallest resolved haloes. Results presented in
Sec.~\ref{Sec:Force} indicate that this force resolution is sufficient
for accurate estimates of $\Vmax$.

Figure~\ref{fig:TwoBody2} shows the circular velocity profiles in
these simulations with very small number of particles.  We can only
estimate the velocity profile for distances larger than $\sim r_s$:
there is not enough particles to probe smaller radii. However, on
average the models stay close to the initial NFW profile: deviations
are less than 5\% for radii $r=(1-10)r_s$ and for time $t<50\, t_{\rm  cross}$.  

The simulations indicate that the value of $\Vmax$ does not show any
systematic changes over the same period of time and that even with 150
particles one can accurately estimate $\Vmax$. At later moments $\Vmax$
gradually increases presumably due to accumulation of the two-body
scattering effects. However, the change is small: 5\% after 200
crossing times.
\begin{figure}
      \includegraphics[width=84mm]{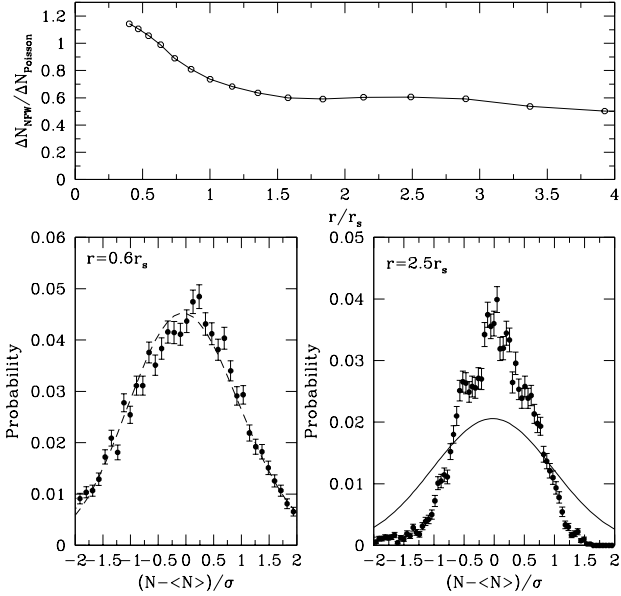}
      \caption{Fluctuations of the number of particles $N$ at
        different distances $r$ from the halo center. Bottom panels
        show the distributions for the central region $0.6\, r_s$ (left
        panel) and for the region $2.5\, r_s$ (right panel), which is
        close to the radius defining the maximum circular velocity
        $\Vmax$. The horizontal axes show deviations from the average
        number of particles $\langle N\rangle$ in units of the r.m.s.
        deviations expected for the Poisson distribution. Curves
        show predictions for the Poisson noise and the points with
        error bars present measurements in $N-$body simulation. The
        top panel shows the r.m.s. fluctuations of the number of
        particles at different radii expressed in units of the r.m.s.
        fluctuations in the Poissonian distribution with the same
        $\langle N\rangle$. The fluctuations in the NFW halo are about
        twice smaller at large distances and become Poissonian at the
        center.}
\label{fig:Distribution}
\end{figure}

In order to measure the level of statistical fluctuations of $\Vmax$,
the value of $\Vmax$ is estimated for each snapshot and the statistics is
accumulated for many snapshots.  Results are presented in the top
panel of Figure~\ref{fig:TwoBody2}. The level of fluctuations of
$\Vmax$ is just $\sim$3\%, which is remarkably small considering that there
are only $N\sim$50 particles inside the radius $r_{\rm max}=2.16\, r_s$, which typically
defines the maximum of the circular velocity. One would naively expect
larger fluctuations of $\sim 1/\sqrt{N}\approx 15\%$. The real fluctuations
are about five times smaller.

  Assuming that the number of particles $N$ inside a given radius $r$
  is distributed according to the Poissonian statistics, we expect
  that the r.m.s. fluctuations of circular velocity are $\Delta V/V =
  \sqrt{\Delta M/M} = \sqrt{\Delta N/N}$, which gives $\Delta V/V
  =1/2\sqrt{N}$. Additional reduction in the fluctuations is related with
  deviations from the Poissonian noise. Fluctuations of the number of
  particles are actually not expected to be Poissonian. One can design
  few examples to illustrate this. Orbital eccentricity should be a
  factor: there are no fluctuations, if the orbits are circular. For a
  given radius there are always some number of particles, which do not
  have energy to leave the radius. These particles always stay inside
  the radius and, thus, their number does not fluctuate. The fraction of
  these particles depends on the radius. In the central region, $r<r_s$,
  a large fraction of particles is not bound to the center. They
  travel to large distances and only occasionally are found within the
  radius $r$. At large distances the fraction of bound particles
  increases resulting in smaller fluctuations.

  To measure the deviations from the Poissonian statistics, we run a
  number of different numerical experiments trying to estimate the r.m.s.
  fluctuations at different radii. The results, when expressed in
  units of the Poissonian noise, do not depend on how many particles
  are used for a particular simulation. Figure~\ref{fig:Distribution}
  shows the results for a simulation with 2,000 particles, which was run
  for 200 crossing-times. The fluctuations in the number of particles
  are close to  Poissonian for small radii. At radius $r=(2-3)r_s$
  the r.m.s. fluctuations are about 1/2 of those expected for the
  Poissonian distribution. Combining all the effects, we estimate that
  the r.m.s. fluctuations of the maximum circular velocity $\Vmax$ due to
  uncorrelated statistical fluctuations of the number of particles can be estimated as:

\begin{equation}
     \frac{\Delta\Vmax}{\Vmax} = \frac{1}{4\sqrt{N}},  
\end{equation}
where $N$ is the number of particles within a sphere of radius $2\,
r_s$. For $N=50$ this approximation gives $\Delta V/V=0.035$, which is
what we found in our simulations.
\begin{figure}
      \includegraphics[width=84mm]{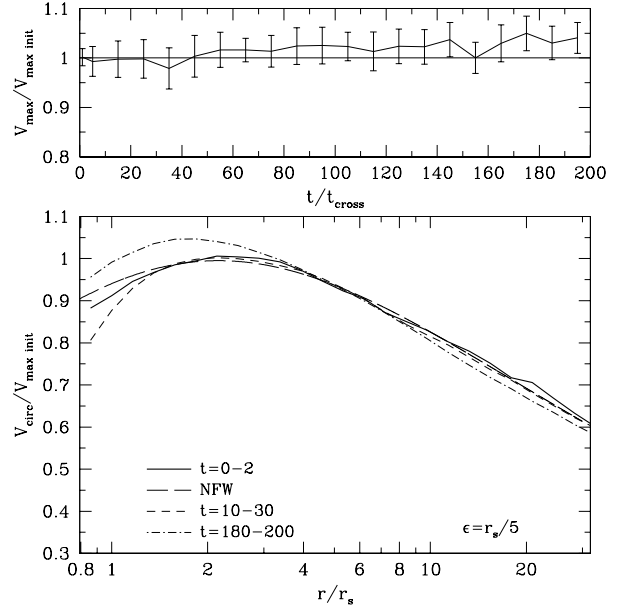}
      \caption{{\it Bottom panel:} Average circular velocity profile
        for haloes with small number of particles at different moments
        of time. The haloes have 25 particles inside $r_s$ and 150
        particles inside the fiducial virial radius of $8\, r_s$.  The
        long-dashed curve shows the NFW profile. Very little evolution
        is observed for the first 50 crossing times, which is close to
        the age of the Universe for haloes in cosmological
        simulations. At later moments effects of the two-body
        scattering gradually modify the profile resulting in slightly
        more dense core and more extended peripheral halo. Time for
        each curve is given in units of the crossing-time $t_{\rm
          cross}$ at the radius $r_{\rm max}=2.16\,r_s$.  {\it Top panel:}
        Average instantaneous maximum of the circular velocity and its
        r.m.s. fluctuations at different moments of time. The
        r.m.s. fluctuations of $\Vmax$ are $\sim 3\%$. }
\label{fig:TwoBody2}
\end{figure}

\subsubsection{Two-body scattering}
Effects of two-body scattering for the NFW haloes were extensively
studied for haloes with relatively large number of particles
\citep{Hayashi2003,Power2003,Valenzuela2003,Diemand2004}.  The
smallest haloes studied had  $\sim 3,000-4,000$ particles
\citep{Hayashi2003,Diemand2004}.  The main focus of these studies was
the inner cusp of the haloes: to what degree the scattering can affect
the inner structure of haloes. Instead, our analysis is focused on very
small haloes with $\sim 100$ particles. Our main interest is
the stability of the region which defines $\Vmax$. All haloes -- even
the largest ones -- go through a stage when they have very few particles.
Hence,  it is important that two-body scattering does not
affect the density of these small haloes. 
\begin{figure}
      \includegraphics[width=84mm]{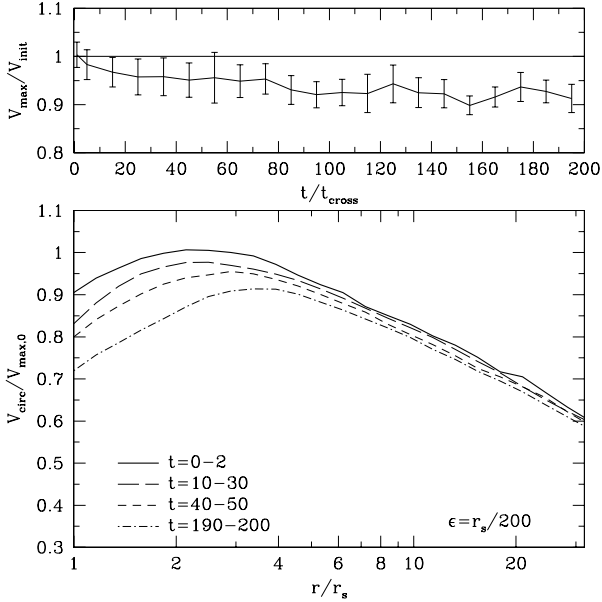}
      \caption{The same as in Figure~\ref{fig:TwoBody2}, but for much
        finer force resolution of $\epsilon=0.005\, r_s$. Two-body
        scattering is much more pronounced as compared with large
        $\epsilon$ simulations typically used in cosmological
        simulations. Even for unrealistically small
        $\epsilon=0.005\, r_s$, the scattering changes $\Vmax$ only by 5\%
        for the life-span of 30--50~$t_{\rm cross}$ of these haloes in
        cosmological runs.}
\label{fig:TwoBody1}
\end{figure}

We start with analytical estimates of the two-body scattering, which
we write using the notations introduced before.  The time-scale $t_{\rm
  relax}$ for the scattering is \citep[][ equation~(7.106)]{BT}:
\begin{equation}
   t_{\rm  relax} =0.34\frac{\sigma^3}{G^2m\rho\ln{\Lambda}},
   \label{eq:trelax}
\end{equation}
where $\sigma$ is the 1d r.m.s. velocity of particles, $m$ is the
particle mass, and $\ln{\Lambda}$ is the Coulomb logarithm.
For the NFW haloes with isotropic velocities one can use the Jeans equation to find $\sigma$:
\begin{equation}
   \left[\frac{\sigma(r)}{\Vmax}\right]^2 = \frac{x_{\rm max}}{f(x_{\rm max})}x(1+x)^2
               \int^\infty_x\frac{f(x)dx}{x^3(1+x)^2}.
   \label{eq:sigma}
\end{equation}
The particle mass $m$ can be parametrized using the number of particles
$N_{\rm max}$ inside the radius $r_{\rm max}=2.16\, r_s$.  Combining these
expressions and writing the two-body scattering time in units of the
crossing time (eqs.\ref{eq:tcross} and \ref{eq:cross}), we get:

\begin{equation}
    \frac{t_{\rm  relax}}{t_{\rm cross}} = 0.34\frac{4\pi N_{\rm max}}{\ln{\Lambda}} 
          \frac{\left[x(1+x)^2\right]^{5/2}}{x^{3/2}_{\rm max}f^{1/2}(x_{\rm max})}
          \left[\int^\infty_x\frac{f(x)dx}{x^3(1+x)^2}\right]^{3/2}. 
   \label{eq:tt}
\end{equation}
The integral in equation~(\ref{eq:tt}) may be taken numerically. However, it can
be approximated in a simple way. We note that the relaxation time
depends on the course-grained phase-space density $\rho/\sigma^3$,
which is know to be nearly a power law. Indeed, the following
approximation is accurate within 15\% for a wide range of radii
$x=0.003-100$:
\begin{equation}
         \left[x(1+x)^2\right]^{5/2} \left[\int^\infty_x\frac{f(x)dx}{x^3(1+x)^2}\right]^{3/2}
          \approx \frac{x^{1.92}}{7.5}.
   \label{eq:apr}
\end{equation}
Substituting this into equation~(\ref{eq:tt}) and estimating numerical factors, we get:
\begin{equation}
    \frac{t_{\rm  relax}}{t_{\rm cross}} = 0.262\frac{N_{\rm max}}{\ln{\Lambda}}\left(\frac{r}{r_s}\right)^{1.92}.  
   \label{eq:tfinal}
\end{equation}
The term $\Lambda$ is the ratio of the maximum $b_{\rm max}$ to
the minimum $b_{\rm min}$ impact parameters for the two-body scattering
problem. In realistic cosmological simulations and in our simple
numerical models the minimum impact parameter is defined by the force
resolution. For $b_{\rm min}$ we use the distance at which the force
becomes Newtonian: $b_{\rm min}=2.8\, \epsilon$. For $b_{\rm max}$ one
may use the whole region, which is relevant for the inner structure of the
NFW distribution $b_{\rm max}\approx (3-10)r_s$.  This leads to
$\ln{\Lambda}\approx \ln(3r_s/\epsilon)$. 

We can get another useful approximation for the relaxation time, if in
equation~\ref{eq:tfinal} we replace $t_{\rm cross}$ with its value
given by equation~(\ref{eq:cross}) and also use the number of
particles inside a given radius: $N(x)=N_{\rm max}f(x)/f(x_{\rm
  max})$.  The following aprroximation gives $t_{\rm cross}$ for a
given halo concentration $c$ and a number of particles $N(r)$:

\begin{equation}
   t_{\rm  relax} \approx 1.4h^{-1}{\rm Gyrs}\frac{N(r)}{c\ln(3r_s/\epsilon)}
            \left[1+\left(\frac{r}{r_s}\right)^{5/4} \right].
   \label{eq:tgyrs}
\end{equation}

%Considering the
%uncertainties, we use $\ln{\Lambda}\approx 2$ for typical force
%resolution $\epsilon = (0.1-0.4)r_s$.

\begin{figure}
      \includegraphics[width=84mm]{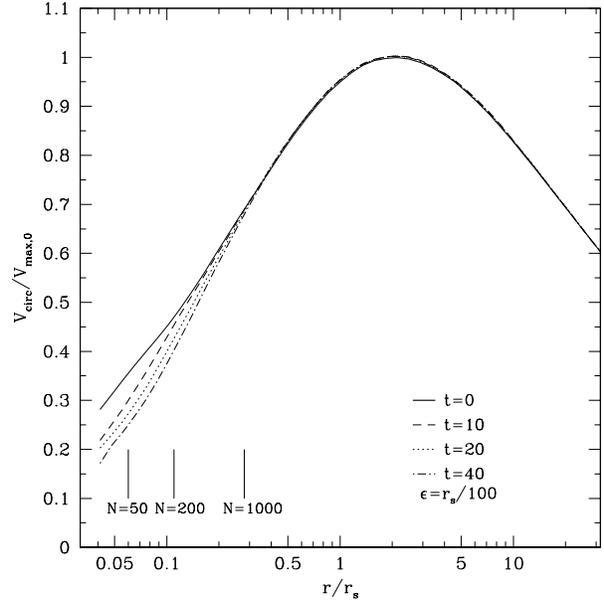}
      \caption{The same as botom panel in Figure~\ref{fig:TwoBody2}, but for
        simulations with 50,000 particles inside  the fidutial virial
        radius. Vertical lines show radii with 50, 200 and 1,000
        particles. Two-body scattering is clearly seen at small
        distances. }
\label{fig:TwoBody3}
\end{figure}

We can use equations~(\ref{eq:tfinal}) or (\ref{eq:tgyrs})  to estimate $t_{\rm relax}$ for our
simulations. For simulations presented in Figure~\ref{fig:TwoBody2},
$N_{\rm max}=45$ and one finds $t_{\rm relax}=26\,t_{\rm cross}$, which is
$10h^{-1}$~Gyrs when scaled to Milky Way-size haloes ($c=10$).

Note that during $t\approx t_{\rm relax}$ the halo density profile
does not change much: within $\sim 5$\% the circular velocity profile
is the same as the initial NFW (see Figure~(ref{fig:TwoBody2})). The
halo does suffer from the scattering, but on much longer time-scale of
$t\approx (5-6)\,t_{\rm relax}$. To some degree, the scattering in
these simulations was suppressed by choosing a reasonable force
resolution, i.e.  small enough to resolve well the halo and not too
small to avoid excessive scattering. Indeed, we run the same
simulations with much smaller force softening and find that scattering
significantly increases. Figure~\ref{fig:TwoBody1} shows the results
for $\epsilon=0.005\,r_s$. It is clear that in this case the two-body
scattering affects the whole halo.  However, even for this unrealistic
case the scattering changes $\Vmax$ by 5\% for the life-span of
$(30-50)\,t_{\rm cross}$ of these haloes in cosmological runs.
\begin{figure}
      \includegraphics[width=84mm]{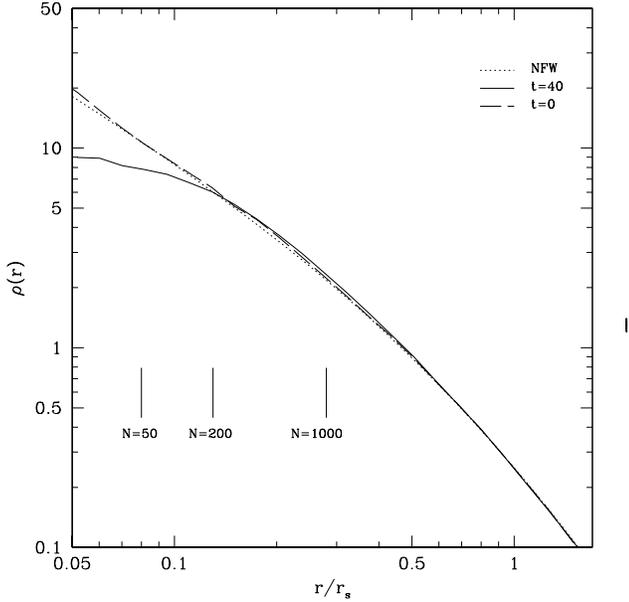}
      \caption{The density profile in the inner region of the NFW
        halo. The full curve shows the density profile after
        40~$t_{\rm cross}$. The other curves show the initial profile
        averaged for $t=0-2t_{\rm cross}$  (long-dashed cirve) and
        the NFW profile (dotted curve). Vertical lines show the
        radii encompassing 50, 200, and 1,000 particles at the final
        stage of evolution. The two-body scattering results in the
        decline of density in the very center. At the radius
        encompasing 200 particles the decline is less than 3\%. }
\label{fig:TwoBody4}
\end{figure}

As equation~(\ref{eq:tfinal}) shows, the relaxation time-scale very
strongly depends on radius $t_{\rm relax}(r)\propto r^{-\alpha},$
$\alpha\approx -2$. This makes the averaging of $t_{\rm relax}$ over
the whole halo not very useful for haloes with steep cusps
\citep{Quinlan,Hayashi2003}. E.g., a long time-scale of scattering
estimated at  half-mass radius does not imply that the scattering is
negligable because it still may be important close to the halo
center. The opposite is also not true: the fact that the scattering
time is short at some distance does not mean that the other parts of the
halo are affected.

All the results present above indicate that the effects of two-body scattering are
localized, with the outer halo regions being basically collisionless.  This
distance, at which the scattering starts to make an effect may or may
not be important for a particular situation.  To illustrate the point we
made numerical experiments with larger number of particles, which allow us to
probe small radii. We simulated three realizations of haloes with 50,000
particles inside the fiducial virial radius of $8r_s$ and $10^5$
particles in total. The force resolution was
$\epsilon=0.01\,r_s$. Figure~\ref{fig:TwoBody3} shows that in these
haloes the central region $r<0.2r_s$ is affected by the two-body
scattering while the outer regions are not. As time goes on, the
radius separating the inner affected region from our collisionless halo
gradually increases.

Note that at the radius containing the first 50 particles the decline
in circular velocity is worse for these haloes with larger number of
particles as compared with the previously studied haloes with only 150
particles inside the fiducial virial radius. This increase in the
scattering at small radii can be shown analytically using
equation~(\ref{eq:tgyrs}). Indeed, the number of particles inside
radius $x$ with a given value of $t_{\rm relax}$ must be $\sim 4$
times larger for $r\ll r_s$ than for $r\approx r_{\rm max}=2.16\,
r_s$.  In that sense, the two-body scattering has the smallest impact
for small haloes. For example, to have $t_{\rm relax}=30\,t_{\rm
  cross}$ a halo should have $\sim 50$ particles inside $x=x_{\rm
  max}$ and $\sim 200$ particles inside $x=0.01$.

Our results are in overall agreement with those of \citet{Hayashi2003}
when we compare their simulation of a NFW halo with 3,000 particles and
take into account differences in the  definitions of the crossing-time and
the Coulomb logarithm. Just as \citet{Hayashi2003}, we also find that
two-body scattering starts to modify the resolved density profile on a
time-scale, which is many times longer than the formal analytical
estimates.  Our conclusions are in strong disagreement with those of
\citet{Diemand2004}, who argue that the two-body scattering so much
affects small haloes that the whole hierarchical growth of haloes is
collisional. This strictly contradicts our results: the two-body
scattering does not affect much the value of $\Vmax$ and the resolved
parts of small haloes. A combination of two factors significantly reduces the
role of the scattering: (a) small life-time of cosmological haloes
($\sim 20-60\,t_{\rm cross}$), and (b) low force
resolution of cosmological simulations.

Our numerical results are also in a very good agreement with those of
\citet{Power2003}, but we disagree on the interpretation and final
conclusions. Results presented in Figure~\ref{fig:TwoBody3} show that
the circular velocity deviates from the analytical solution by less
than $\sim 2$\% for central region containing 1,000 particles. This is
in agreement with Figure~14 in \citet{Power2003}, which shows the mean
halo density contrast as a function of the enclosed number of
particles. However, this is somewhat misleading because both
statistics (circular velocity and mean enclosed density) are integral
characteristics. The plot of the density at a given radius (a differential
statistics) presented in Figure~\ref{fig:TwoBody4} shows that the
density is not affected by the scattering already at radius containing
$\sim 200$ particles. This agrees with the analysis of
convergence of density profiles in cosmological simulations presented
by \citet{Klypin2001}.

It is difficult to justify an analytical approximation for the
two-body scattering time used by \citet[][ equation (20)]{Power2003}. It
gives the impression that it is a slightly modified form of the
Chandrasekhar formula, but it is actually not. At the end the main
difference is related with the r.m.s. velocities. In the Chandrasekhar
approximation for two-body scattering (see equation~(\ref{eq:tt})) the term
$\sigma(r)$ is the r.m.s. velocity of the particles, which in the inner
halo regions  is substantially larger than the circular velocity
$V(r)$ given by equation~(\ref{eq:Vnfw}).  Instead, \citet[][
equation (20)]{Power2003} use the circular velocity. This results in
a substantial overestimate of the two-body scattering time and in
a different scaling relation. One may treat \citet[][
equation (20)]{Power2003} as a pure numerical fit to simulation data, which
is not related to dynamical arguments. However, after correcting for
the integral nature of the data presented in \citet[][
equation (20)]{Power2003}, a better approximation to the data is a simpler
form of the radius contaning $\sim 200$~ particles, which we find in our
simulations.

To summarize, we find that two-body scattering has less than 2\%
effect on the value of $\Vmax$ for haloes with as little as $\sim 50$
particles inside the radius defining $\Vmax$. In order to achieve the same
accuracy for smaller radii $r<r_s$, the number of particles should be
$\sim 200$.

\subsection{Subhaloes: tidal stripping and limits of survival}
When a dark matter  halo falls into another halo and becomes a subhalo, the tidal
forces of the ``parent'' start to remove the least bound particles
from the subhalo, thus reducing its mass.  Effects of the tidal
stripping experienced by an orbiting dark matter halo with a NFW
profile have been extensively studied in the past
\citep[e.g.,][]{Klypin99,Hayashi2003,Penarrubia2008,Arraki}. Tidal
stripping depends on the pericenter of the subhalo orbit, on the
concentration of the subhalo, and on the number of orbits which the
subhalo makes after falling into the parent halo. So, the situation is
complex and it is difficult to find a realistic description of the
whole problem without running complicated simulations. Nevertheless,
we can derive simple relations, which can be used to understand
the survival of subhaloes in our cosmological simulations. The main
question, which we try to answer here, is how can a small halo with
only $\sim 100$ particles survive when it falls into a larger halo with
a strong tidal force?

To be more specific, we provide here simple estimates of the minimum
distance from the center of the parent at which the subhalo, with a
given number of particles, can be detected in cosmological
simulations. Our analysis is simplified by the fact that the decrease
in the mass of the central region of the halo with $r\lesssim r_s$ is
very sensitive to the ratio between the tidal radius $r_{\rm tide}$
and the scale radius $r_s$ of the NFW density profile. For $r_{\rm
  tide}> 3\,r_s$ the central density and $\Vmax$ do not change much. For
example, according to Figure~8 in \citet{Arraki} for $r_{\rm tide} =
3\,r_s$ the value of $\Vmax$ declines just by 20-30\% as compared with
the initial NFW value. This subhalo potentially can be detected by a
halofinder. The situation is quite different for a smaller tidal
radius. For example, for $r_{\rm tide} = 2\,r_s$ the maximum circular
velocity decreases to 1/3-1/2 of its initial value. This means that
the mass of the object becomes less than $\sim 20-30$ particles and
subhaloes with this few particles cannot be detected.

The reason for this steep decline of bound mass is related to the fact
that the central $r\lesssim r_s$ region is not self bound: kinetic
energy of all particles inside this region is larger than the potential
energy of these particles. Once the tidal radius is close to this
unbound region, tidal stripping dramatically increases. It does
not mean that the whole subhalo is totally disrupted because a small
central region still may survive \citep{Penarrubia2008}. However, in
simulations with relatively small number of particles, the subhalo is
lost. This is the reason why we assume that a subhalo is destroyed and
cannot be detected in simulations once the tidal radius becomes less
than $\sim 2\, r_s$. Our goal is to find the distance $R_{\rm lim}$
to center of the parent halo at which this happens.

There are two conditions which are used to find the tidal radius
$r_{\rm tide}$ \citep[e.g.,][]{Klypin99}. The first is found by
equating the external tidal force to the force due to the interior of
the subhalo. The second is related with the resonance between the
external force and the internal orbital motion of particles.  We use
the resonant condition for orbital frequencies, which gives slightly
smaller tidal radius in central halo regions. If $\omega(r)$ is the
frequency of a particle on a circular orbit with radius $r$ from the
center of the subhaloes and $\Omega(R)$ is the frequency of the
perturbing external force, then the tidal radius $r_{\rm tide}$ is
defined by condition $\omega(r_{\rm tide})=\Omega(R)$.

The ratio of orbital frequencies  can be written in the following form:
\begin{equation}
    \frac{\omega^2(r)}{\Omega^2(R)} = \left(\frac{V_{\rm max}}{V_{\rm p,max}}\right)^2
          \left( \frac{R_{s}}{r_s}  \right)^2
           \left(\frac{y}{x}  \right)^3
          \frac{f(x)}{f(y)}, \quad y\equiv \frac{R}{R_s},
  \label{eq:frratio}
\end{equation}
where $V_{\rm p,max}$ and $R_s$ are the maximum circular velocity and
the characteristic radius of the parent halo.  Equation~(\ref{eq:frratio})
can be simplified if we use halo concentrations $c_p$ and $c_s$ for
the parent and the subhalo correspondingly. In this case, the equation
for the tidal radius $x_{\rm tide}=r_{\rm tide}/r_s$ takes a simple
form:
\begin{equation}
  u(x_{\rm tide}) =\frac{u(c_s)}{u(c_p)}u(y), \qquad u(x) = \frac{x^3}{f(x)}.
  \label{eq:utide}
\end{equation}
Note that function $u(x)$ has a simple meaning. It is inverse of the
mean density inside radius $x$.  Equation~(\ref{eq:utide}) can be
solved numerically to find the tidal radius $x_{\rm tide}$ for given
halo concentration pair $c_p, c_s$ and for the distance from the
parent center $y$. It also can be used to find the distance $y_{\rm
  lim}$ from the parent at which the subhalo is destroyed. To find
$y_{\rm lim}$ we assume that the subhalo is destroyed when its tidal
radius $x_{\rm tide}$ becomes too small. If $x_d$ is the radius, than
$y_{\rm lim}$ can be found by solving the following equation
\begin{equation}
  u(y_{\rm lim}) =\frac{u(c_p)}{u(c_s)}u(x_d).
  \label{eq:xtide}
\end{equation}
 
The following approximations give 15\% accuracy for the tidal radius
and for the destruction distance:
\begin{eqnarray}
  r_{\rm tide} &\approx& \frac{R}{1+0.04(R/R_s)^{3/2}} \left(\frac{r_s}{R_s}\right)
\left(\frac{c_s}{c_p}\right)^{3/2},   \label{eq:rtide} \\
  R_{\rm lim} &\approx& R_s\frac{0.9x_d}{1+0.04x_d^{3/2}}\left(\frac{c_p}{c_s}\right)^{3/2}, \, x_d=0.1-3. 
  \label{eq:Rlim}
\end{eqnarray}
One interesting consequence of
equations~(\ref{eq:rtide}-\ref{eq:Rlim}) is that in a typical
large-scale cosmological simulation subhaloes should be destroyed if
their distance $R$ to the parent becomes less than the scale radius
$R_s$ of the parent.  Indeed, unless an exuberant number of particles
is used for a subhalo, the subhalo will be destroyed once the tidal
radius becomes less than $(1-2)\,r_s$. Denser subhaloes survive better
against the tidal field: the tidal radius $ r_{\rm tide}\propto
c_s^{3/2}$. However, the average concentration depends very weakly on
mass $c(m)\propto m^{-0.1}$.  For subhaloes which are 100-1000 times
less massive than the parent halo, $c_s/c_p \approx 1.5-2$.  Putting
these estimates into equation~(\ref{eq:Rlim}), we find that a
combination of tidal stripping and small number of particles inside
subhalos make survival of subhalos difficult for distances from the
center of parent halo smaller than $R_{\rm lim}=(0.3-0.9)R_s$.

\section{Conclusions}
\label{sec:Conclusions}

Large cosmological $N$-body simulations, which use billions of
particles and provide properties of millions of dark matter haloes,
play a very important role in testing predictions of the standard
cosmological model. It takes significant amount of computer resources
and manpower to run and analyze these simulations. In this respect, it
is crucial to optimize numerical parameters used to make the
simulations and to understand their limitations. There is no unique
set of requirements for the simulations: it all depends how the
simulations will be used for  particular science applications.  For
numerous purposes resolving the interior structure of haloes and
identifying subhaloes are needed.  Using state-of-the-art cosmological
simulations and simplified models of individual dark matter haloes, we
investigate the numerical accuracy and convergence of (sub)halo
properties used for the Halo Abundance Matching method.

In the large volume cosmological simulations with billions of particles,
which use the $N$-body codes ART and Gadget, we find that convergence
at the 10\% accuracy for the abundance of haloes and subhaloes, and the
correlation functions can be achieved with $\sim 150$ particles when
parameters of the simulations -- the force resolution and time-stepping --
are chosen to satisfy a number of conditions. 

Using simplified models of dark matter haloes, we study different
effects, which may play a key role in the accuracy of results. Force
resolution is the key parameter.  Figure~\ref{fig:ForceResolution}
shows how the circular velocity profile is affected by the lack of
force resolution. Equation~(\ref{eq:Vcorrect}) can be used to estimate
the error of $\Vmax$ for a given force resolution $\epsilon$. Because
the error in $\Vmax$ produces much larger error in the velocity
function ($n(\Vmax)\propto \Vmax^{-3}$), we recommend that  the force
resolution (equivalent of the Plummer softening) should be
\begin{equation}
 0.1r_s <  \epsilon<0.3r_s,
\label{eq:ForcR}
\end{equation}
where $r_s$ is the scale radius of smallest resolved haloes. Here the
upper limit is defined by condition that the error in $\Vmax$ is less
than 5\%. At the lower limit the error is 1\%. The lower limit is set
to avoid too fine resolution, which may enhance the two-body
scattering and may require too small time-step.  

These constraints on the force resolution should be contrasted with
the ``optimal gravitational softening'' recommended by
\citet{Power2003}: $\epsilon/r_s > 4c/\sqrt{N}$, where $c$ is the halo
concentration and $N$ is the number of particles inside virial
radius. Assuming $c=10$ and $N=150$, we get $\epsilon/r_s >3.2$. This
force softening would result in errors in $\Vmax$ so large that it
would render the simulation useless. Note that our recommended upper
limit on $\epsilon$ is ten times smaller than the lower limit in
\citet{Power2003}. 

We also find that the two-body scattering, though clearly detected, plays
relatively minor role for small haloes with as little as $\sim 100$
particles. Estimates based on standard analytical approximations such
as equation~(\ref{eq:trelax}) and equation~(\ref{eq:tfinal}) are still useful
because they provide insights on scaling of the relaxation time
with distance and number of particles. However, analytical approximations
overestimate the effect of the scattering.  For the region which defines the
value of the maximum circular velocity $\Vmax$ the analytical
estimates give relaxation times, which are 5--6 times too short. 

Two effects seem to contribute to the reduced impact of the two-body
scattering in cosmological haloes: small number of crossing times for
particles and reduced force resolution.

Time-stepping plays an important role in defining the accuracy of
simulations \citep[e.g.,][]{Power2003,Klypin2009}. However, because of
different implementations of time-stepping schemes and conditions for
time-step refinement, it is difficult to give unique
recommendations. For our Gadget simulations we use parameter {\it
  ErrTolIntAccuracy}$=0.01$, which is twice smaller than what was used
for the Millennium-I simulation. The time-step was even smaller in the
case of MultiDark, which used $\sim 40,0000$ steps at the highest
level of resolution.

Tidal stripping and numerical destruction of subhaloes also set
limitations on the HAM models. As equation~(\ref{eq:Rlim}) shows, in
large cosmological simulations subhaloes are destroyed once they come
too close to the parent's center. For realistic combinations of halo
concentrations, this subhalo destruction occurs at the distance from
the parent halo $R_{\rm lim} \approx (0.3-1) R_s$. Subhaloes can
survive at smaller distances, if they have a very large number of
particles. They also can be found there temporarily because it takes
few orbits to get severely stripped, which can take few billion years.

We recommend the following steps to estimate parameters of $N$-body
simulations, which can be used to resolve central halo regions that
define (sub)halo maximum circular velocity:

\begin{enumerate}
\item Using particle mass $m$ find virial mass $\Mvir$ containing
  $150$ particles. Find the halo virial radius. Use a concentration -
  mass relation to find halo concentration $c$ for smallest resolved
  halos.

\item Using the halo concentration $c$ and the virial radius, find scale
  radius $r_s$ for the smallest halo. Plummer force softening $\epsilon$
  is defined by required accuracy of the maximum circular velocity
  $\Vmax$. Use equation~(\ref{eq:Vcorrect}) to find $\epsilon$, which
  gives this accuracy.

\item Run small tests with these parameters to find the time-step (or
  parameters, which define it) that give converge of halo profiles and
  (sub)halo abundances. The smaller is the force softening, the
  smaller is the time step.

\end{enumerate}
\section*{Acknowledgments}

AK acknowledges the support of NSF and NASA grants to NMSU.  GY
acknowledges support from the Spanish MINECO under research grants
AYA2012-31101, FPA2012-34694 and Consolider Ingenio SyeC CSD2007-0050
and from Comunidad de Madrid under ASTROMADRID project
(S2009/ESP-1496). The BigMultiDark simulation suite have been
performed in the Supermuc supercomputer at LRZ thanks to the cpu time
awarded by PRACE (proposal number 2012060963). 
S.H. acknowledges support by the Deutsche 
Forschungsgemeinschaft under the grant $\mathrm{GO}563/21-1$. 

 \bibliography{Comparison} \bibliographystyle{apj}

\end{document}